\documentclass[a4paper,11pt]{article}
\usepackage{pos}

\usepackage[symbol]{footmisc}
\usepackage[english]{babel}
\usepackage{graphicx}
\usepackage{graphics}
\usepackage{braket}
\usepackage{bbold}
\usepackage{amsmath}
\usepackage{nicefrac}
\usepackage{dcolumn}
\usepackage{bm}
\usepackage{slashed}
\usepackage{datetime}
\usepackage{mciteplus}
\usepackage{multirow}
\usepackage{siunitx}
\usepackage{booktabs}
\usepackage{color, soul}
\usepackage[usenames,dvipsnames]{xcolor}
\usepackage{float}
\usepackage[utf8]{inputenc}
\usepackage[normalem]{ulem}
\usepackage{mathtools}
\usepackage{setspace}
\usepackage{comment}
\renewcommand{\thefootnote}{\fnsymbol{footnote}}

\newcommand{\LQCD}{\Lambda_{\rm QCD}}

\newcommand{\NLLm}{{\rm NLL/NLO^-}}

\newcommand{\DY}{\Delta Y}

\newcommand{{\HFNRevo}}{\tt HF-NRevo}

\title{Angular correlations and polarization in Z+jet production at the LHC}
\ShortTitle{Z+jet production at the LHC}

\author*[a]{Francesco Giovanni Celiberto}
\author[a]{Francesca Lonigro}

\affiliation[a]{Departamento de Física y Matemáticas, Universidad de Alcalá (UAH), Campus Universitario, \\ Alcalá de Henares, E-28805, Madrid, Spain}

\emailAdd{francesco.celiberto@uah.es}
\emailAdd{francesca.lonigro@uah.es}

\abstract{Precision studies at the LHC increasingly require theoretical control over QCD radiation beyond fixed-order perturbation theory, particularly in final states characterised by large separations in rapidity. 
High-energy logarithms can generate corrections at the ten-percent level even for electroweak-scale observables, as recently observed in Higgs production, motivating their systematic inclusion in precision analyses. 
We investigate this regime through Drell--Yan plus jet production, with particular emphasis on $Z$-boson final states and their angular and polarisation structure. 
High-energy logarithms are resummed at next-to-leading logarithmic accuracy and consistently combined with NLO fixed-order information. 
Beyond inclusive rapidity and transverse-momentum spectra, the framework resolves the azimuthal harmonic content of the process, whose pattern is reshaped by the transition from leading- to next-to-leading-order emission dynamics. 
We further develop the JETHAD--DYnamis--POWHEG strategy towards realistic lepton-level predictions, retaining spin correlations, decay-angle information, and experimentally accessible kinematics. 
This programme establishes $Z$+jet production as a precision probe of high-energy QCD dynamics for Run~3 and the HL-LHC.}

\FullConference{14th Edition of the Large Hadron Collider Physics (LHCP2026) \\
18-22 May 2026 \\
Paris, France
}

\begin{document}
\maketitle

\section{High-energy QCD in LHC precision observables}
\label{sec:introduction}

The precision programme of the LHC increasingly explores final states in which several perturbative scales coexist with a much larger hadronic centre-of-mass energy.
The same hierarchy will become even more pronounced at prospective FCC facilities~\cite{FCC:2025lpp,FCC:2025uan,FCC:2025jtd}.
When
$\sqrt{s}\gg{Q}\gg\LQCD$, with $s$ the center-of-mass energy squared, ${Q}$ a set of process-characteristic hard scales, and $\LQCD$ the QCD scale, powers of the strong coupling can be accompanied by sizeable logarithms of the available energy, so that a reliable perturbative description requires their resummation to all orders.
The Balitsky--Fadin--Kuraev--Lipatov (BFKL) formalism~\cite{Fadin:1975cb,Balitsky:1978ic} provides this resummation at leading- and next-to-leading-logarithmic accuracy and, at the same time, gives access to the low-$x$ gluon dynamics of the proton~\cite{Bacchetta:2020vty,Bacchetta:2024fci,Amoroso:2022eow,Bolognino:2018rhb,Bolognino:2021niq,Hentschinski:2022xnd}.
At the LHC, a particularly sensitive configuration is obtained by tagging two hard objects on opposite sides of a sizeable rapidity interval $\DY$.
The transverse scales keep the process perturbative, whereas the large separation enhances real radiation in the intervening rapidity region and exposes the characteristic multi-Regge dynamics.
Within hybrid factorisation (HyF)~\cite{Celiberto:2020tmb,Bolognino:2021mrc}, the corresponding cross section is described by process-dependent emission functions linked through the universal NLL BFKL Green's function.
Full NLL/NLO accuracy is reached when both emission coefficients are available at NLO; when one of them is known only at LO, the mixed $\NLLm$ construction retains NLL evolution while combining NLO and LO information at the two ends of the ladder.
Complementary high-energy approaches to single-inclusive reactions have been developed in~\cite{Bonvini:2018ixe,Silvetti:2022hyc}.
An important feature of this formulation is that the resummation acts not only on inclusive rates but also on the azimuthal structure of the final state.
The cross section can be decomposed into angular harmonics, whose relative weights encode the degree of correlation between the tagged objects.
Radiative emissions progressively decorrelate the final state, while NLO corrections to the emission functions can modify the relative hierarchy of these harmonics.
Angular observables therefore provide a more differential test of high-energy dynamics than total rates alone and are especially valuable when moving towards polarisation-sensitive electroweak final states.
High-energy resummation has already been investigated in a wide range of semi-inclusive channels, including Mueller--Navelet jets~\cite{Ducloue:2013hia,Celiberto:2015yba,Celiberto:2016ygs,Celiberto:2017ius,Celiberto:2022gji}, forward Drell--Yan production~\cite{Celiberto:2018muu,Golec-Biernat:2018kem}, identified hadrons~\cite{Celiberto:2016hae,Celiberto:2022kxx,Celiberto:2017nyx,Bolognino:2019yls,Celiberto:2021dzy,Celiberto:2021fdp,Celiberto:2022zdg,Celiberto:2022keu,Celiberto:2024omj,Feng:2022inv,Celiberto:2022grc}, quarkonia~\cite{Boussarie:2017oae,Celiberto:2022dyf,Celiberto:2023fzz,Celiberto:2022grc}, and rare or exotic final states~\cite{Celiberto:2025ogy,Celiberto:2026qiz,Celiberto:2023rzw,Celiberto:2024mab,Celiberto:2024mrq,Celiberto:2024beg,Celiberto:2025dfe,Celiberto:2025ziy,Celiberto:2026kks,Celiberto:2025ipt,Celiberto:2026ooh}.
Higgs+jet production provides a particularly useful benchmark for bringing this resummation programme into the precision sector.
High-energy calculations at LHC and FCC energies~\cite{Celiberto:2020tmb,Celiberto:2023rtu} can be confronted with fixed-order predictions through NNLO~\cite{Chen:2014gva,Boughezal:2015dra,Dawson:2022zbb} and with next-to-NLL transverse-momentum resummation~\cite{Monni:2019yyr}, thereby testing complementary logarithmic limits within the same process.
Further progress towards higher high-energy accuracy is represented by the recent two-loop Higgs impact factor in the Regge limit~\cite{DelDuca:2025vux}.
This makes Higgs+jet production an ideal environment in which to validate the matching strategy before applying it to more differential electroweak observables.
Our principal LHC target is Drell--Yan plus jet production, with particular attention to a $Z$ boson recoiling against a rapidity-separated jet.
The leptonic decay of the $Z$ provides experimentally reconstructible polarisation and decay-angle information, allowing the high-energy dynamics to be tested through both momentum spectra and angular coefficients.
Within the JETHAD ecosystem, DYnamis represents the dedicated supermodule for Drell--Yan final states, designed to preserve the angular structure, polarisation information, and lepton-level kinematics needed for this programme.

\section{Higgs+jet production as a matching benchmark}
\label{sec:matching}

Earlier studies of Higgs+jet production within HyF revealed a notable perturbative stability at both LHC~\cite{Celiberto:2020tmb} and FCC~\cite{Celiberto:2023rtu} energies.
Nevertheless, a direct comparison with fixed-order calculations showed visible differences in kinematic regions where both descriptions should provide meaningful information.
Rather than treating these predictions as alternative approximations, this motivates their combination within a common perturbative construction.
We adopt an additive matching prescription in which the NLO fixed-order result is supplemented by the genuinely resummed high-energy contribution.
The latter is isolated by expanding the NLL result to fixed order and subtracting this expansion before combining it with the complete NLO calculation.
This procedure removes the perturbative terms common to the two descriptions and therefore prevents double counting while retaining both the exact fixed-order content and the all-order energy logarithms.
At present, the full NLO Higgs emission function~\cite{Hentschinski:2020tbi,Celiberto:2022fgx,Celiberto:2024bfu,Celiberto:2026_Higgs-hadron_NLL-NLO} has not yet been incorporated into JETHAD~\cite{Celiberto:2020wpk,Celiberto:2022rfj,Celiberto:2023fzz,Celiberto:2024mrq,Celiberto:2024swu}.
The phenomenological implementation considered here therefore employs the mixed $\NLLm$ accuracy, where one emission function is taken at NLO and the other at LO while the exchanged Green’s function remains fully NLL.
The POWHEG+JETHAD matching architecture has been developed in Refs.~\cite{Celiberto:2023uuk,Celiberto:2023eba,Celiberto:2023nym,Celiberto:2024mdt} and can be schematically written as

\begin{equation}
\label{eq:matching}
\begin{split}
 \hspace{-0.255cm}
 \underbrace{{\rm d}\sigma^{{{\rm NLL/NLO}}^{\boldsymbol{-}}}(\Delta Y, \varphi, s)}_{\text{\colorbox{OliveGreen}{\textbf{\textcolor{white}{NLL/NLO$^{\boldsymbol{-}}$}}} {\tt POWHEG+JETHAD}}} 
 = 
 \underbrace{{\rm d}\sigma^{\rm NLO}(\Delta Y, \varphi, s)}_{\text{\colorbox{gray}{\textcolor{white}{\textbf{NLO}}} {\tt POWHEG} w/o PS}}
 +\; 
 \underbrace{\underbrace{{\rm d}\sigma^{{{\rm NLL}}^{\boldsymbol{-}}}(\Delta Y, \varphi, s)}_{\text{\colorbox{red}{\textbf{\textcolor{white}{NLL$^{\boldsymbol{-}}$ resum}}} (HyF)}}
 \;-\; 
 \underbrace{\Delta{\rm d}\sigma^{{{\rm NLL/NLO}}^{\boldsymbol{-}}}(\Delta Y, \varphi, s)}_{\text{\colorbox{orange}{\textbf{NLL$^{\boldsymbol{-}}$ expanded}} at NLO}}}_{\text{\colorbox{NavyBlue}{\textbf{\textcolor{white}{NLL$^{\boldsymbol{-}}$}}} {\tt JETHAD} w/o NLO$^{\boldsymbol{-}}$ double counting}}
 \,.
\end{split}
\end{equation}

\begin{figure*}[!t]
\centering

\includegraphics[scale=0.37,clip]{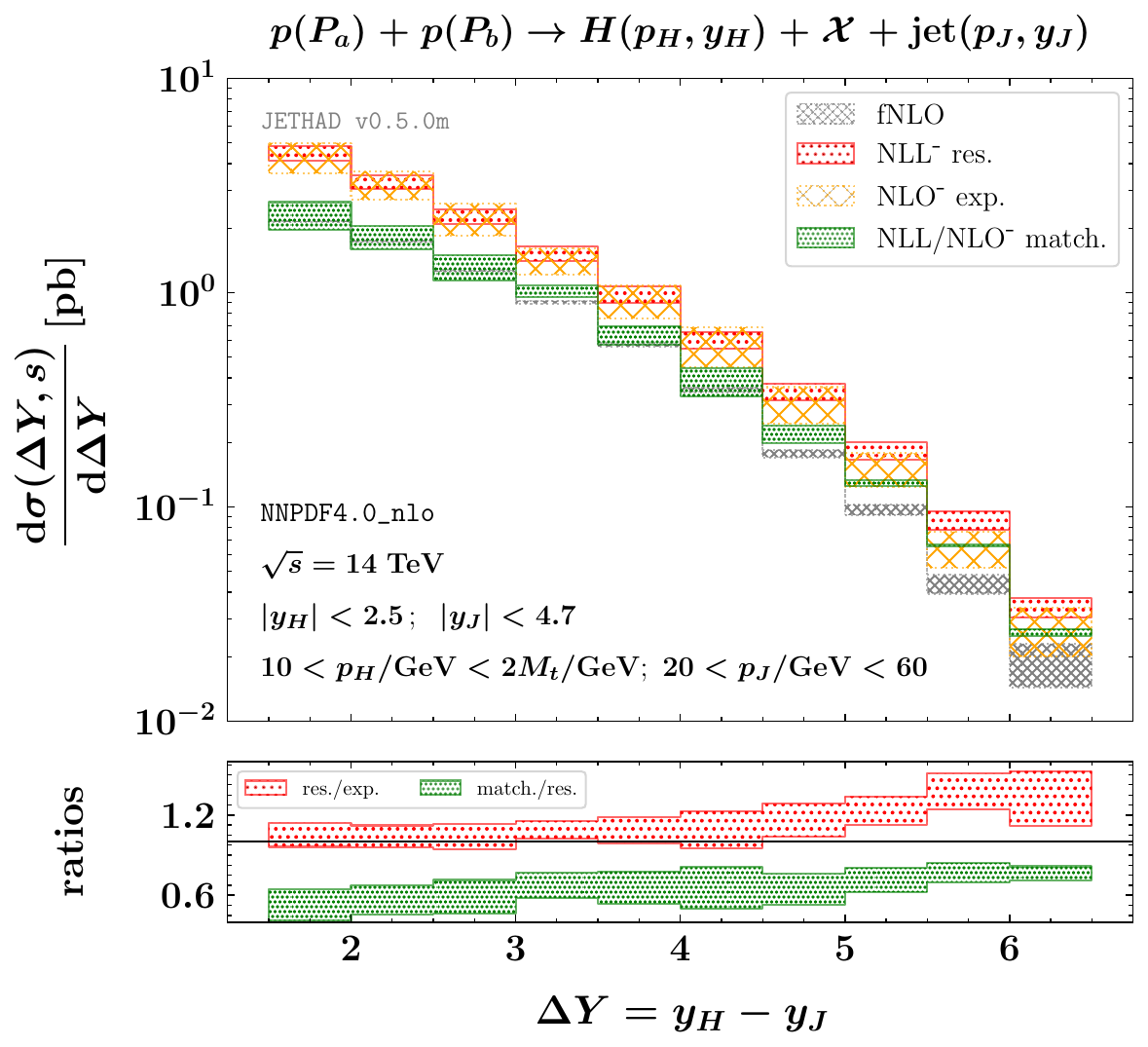}
\hspace{0.00cm}
\includegraphics[scale=0.37,clip]{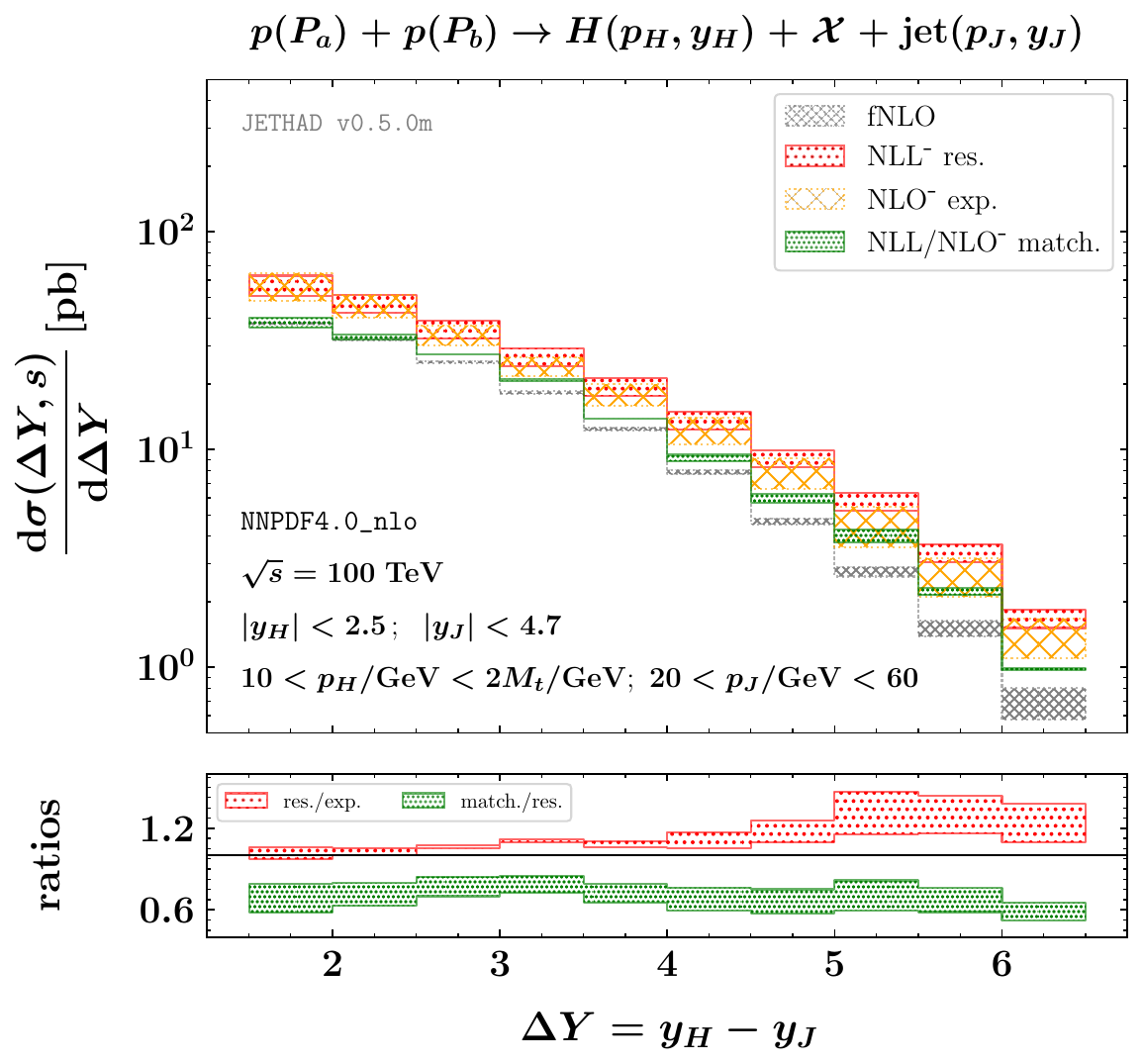}

\caption{Rapidity-separation spectrum for Higgs+jet production at the 14~TeV LHC (left) and the nominal 100~TeV FCC (right), comparing fixed-order, high-energy-resummed, expanded, and matched predictions. 
Bands correspond to $\mu_{R,F}$ scale variations in the range $1<C_{\mu}<2$; the adopted fiducial selections are specified in the text boxes.}

\label{fig:I}
\end{figure*}

\begin{figure*}[!t]
\centering

\includegraphics [scale=0.36,clip]{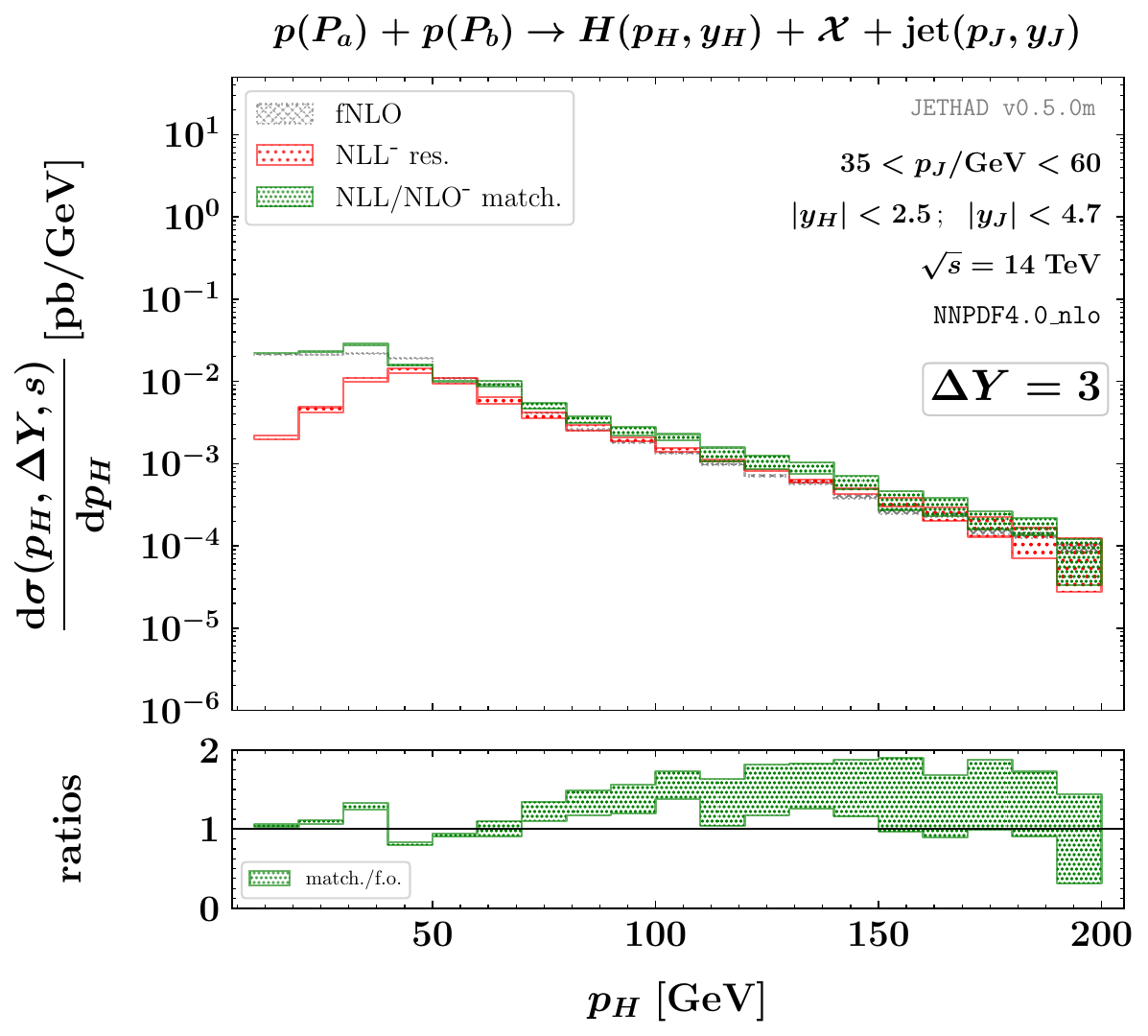}
\hspace{0.00cm}
\includegraphics[scale=0.36,clip]{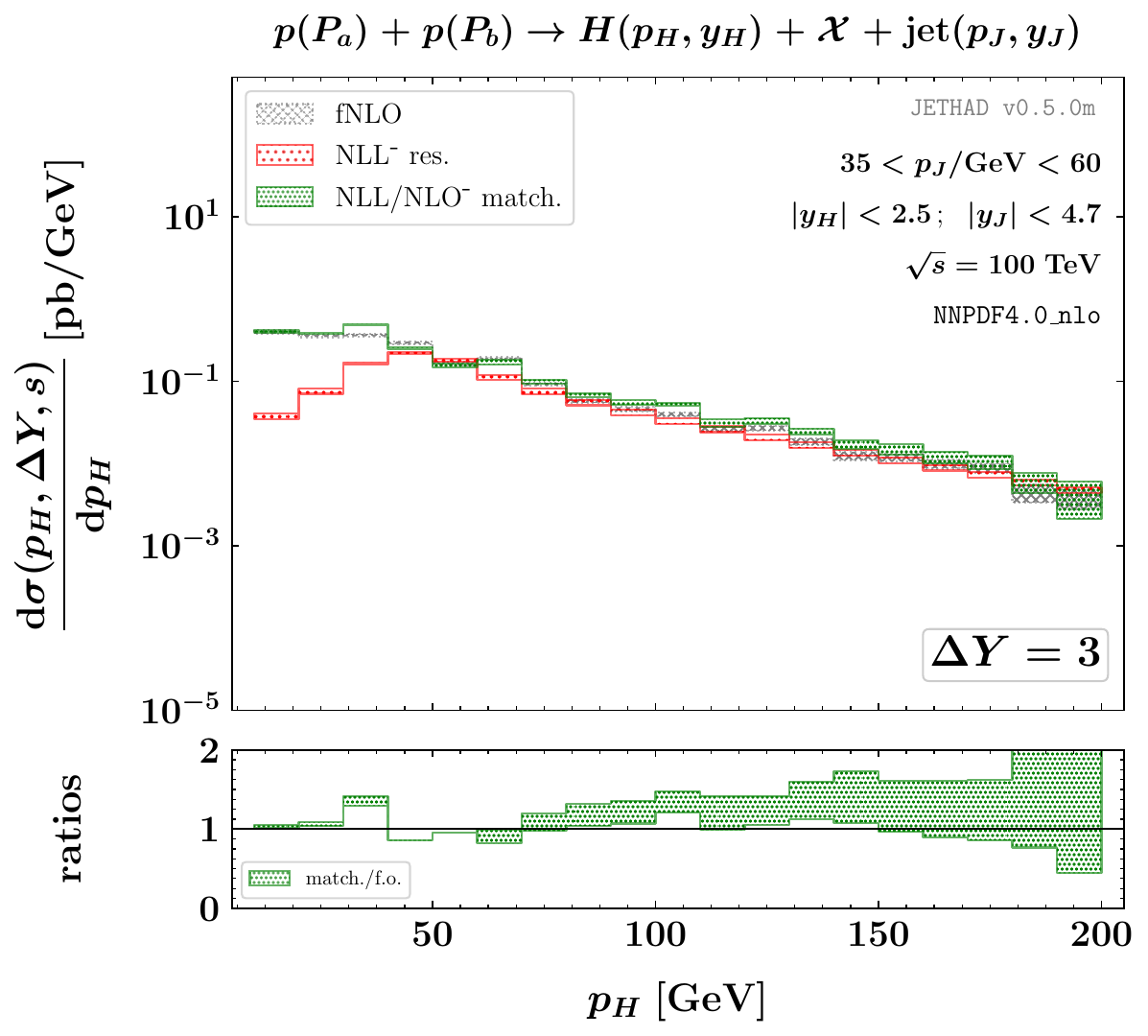}

\caption{Higgs transverse-momentum spectrum in Higgs+jet production at the 14~TeV LHC (left) and the nominal 100~TeV FCC (right). 
The comparison illustrates the impact of high-energy resummation and fixed-order matching around the spectral peak and towards the large-transverse-momentum region.
Bands correspond to $\mu_{R,F}$ scale variations with $1<C_{\mu}<2$.}

\label{fig:pT}
\end{figure*}

The first term in Eq.~\eqref{eq:matching} is the NLO prediction generated with POWHEG~\cite{Hamilton:2012rf,Bagnaschi:2023rbx,Banfi:2023mhz}, evaluated here before the inclusion of parton-shower effects.
The second term represents the net high-energy correction supplied by JETHAD: the NLO truncation of the NLL$^-$ series is removed from the complete resummed result before the latter is added to the fixed-order calculation.
The graphical convention used in Figs.\ref{fig:I} and\ref{fig:pT} mirrors this decomposition, with the fixed-order, resummed, expanded, and matched predictions shown in grey, red, orange, and green, respectively.
Figures~\ref{fig:I} and~\ref{fig:pT} compare this construction between the 14~TeV LHC and the nominal 100~TeV FCC, extending the earlier analyses of Refs.~\cite{Celiberto:2023uuk,Celiberto:2023eba,Celiberto:2023nym}.
The first observable is the rapidity interval $\DY$, whereas the second is the Higgs transverse-momentum distribution at fixed $\DY=3$.
Their ratio panels separately expose the size of the all-order resummation and the modification induced by matching.
The rapidity-interval spectrum displays the characteristic high-energy pattern: the resummed prediction progressively departs from its fixed-order expansion as $\DY$ increases.
This reflects the growing phase space available for BFKL radiation and confirms that the high-energy component becomes more important as the tagged objects are moved further apart.
Correspondingly, the matched result approaches the purely resummed one in the large-$\DY$ region.
At the LHC the transition is particularly transparent, whereas at 100~TeV it is distributed over a wider phase-space domain.
The transverse-momentum spectrum provides a complementary test.
Near its maximum, the matched result differs from the fixed-order prediction by approximately 30–50\%, consistently with the sensitivity to logarithms of the form $\ln(s/p_H^2)$ previously highlighted in Ref.~\cite{Celiberto:2020tmb}.
At larger transverse momentum the uncertainty grows, signalling the increasing importance of logarithmic contributions outside the pure high-energy sector, including collinear enhancements and threshold effects~\cite{Bonciani:2003nt,deFlorian:2005fzc,Muselli:2017bad}.
A more complete treatment of this region will therefore require the interplay of complementary resummation mechanisms.
Most importantly for the present programme, the matched construction reduces the discrepancy between fixed-order and pure-HyF predictions observed in the large-transverse-momentum region, including that reported in Fig.~8 of Ref.~\cite{Celiberto:2020tmb}.
Higgs+jet production thus acts as a controlled validation channel for the POWHEG+JETHAD interface.
The same matching logic can then be transferred to $Z$+jet production, where the additional angular degrees of freedom provide substantially more differential information on the underlying high-energy dynamics.

\section{Towards angular-resolved $Z$+jet phenomenology}
\label{sec:roadmap}

The Higgs benchmark shows that high-energy resummation can produce phenomenologically relevant effects even within the precision sector of LHC physics.
For Drell--Yan production, however, the available information is considerably richer.
Once the intermediate electroweak boson is reconstructed through its leptonic decay, the analysis is no longer restricted to inclusive momentum distributions: the decay-system orientation retains information on the polarisation state of the boson and on the QCD dynamics responsible for its production.
This makes $Z$+jet production a particularly attractive channel for Run~3 and the HL-LHC.
Drell--Yan measurements already serve as standard candles for luminosity studies, PDF constraints, detector calibration, and precision electroweak analyses.
Theoretical effects at the few- to ten-percent level can therefore become relevant at the accuracy targeted by future measurements, especially in corners of phase space characterised by a large rapidity separation between the reconstructed boson and the accompanying jet.
From the high-energy point of view, the angular dependence provides an additional discriminator.
At leading order, the emission vertices generate a characteristic hierarchy among the azimuthal harmonics correlating the $Z$ boson and the jet.
Real and virtual corrections entering at NLO modify both their normalisation and their relative pattern, while the NLL BFKL Green's function further reshapes the correlations through radiation emitted across the rapidity interval.
The LO to NLO transition can therefore be studied not only through changes in the overall rate, but also through the modulation of the angular coefficients and their ratios.
When the $Z$ decay is retained explicitly, this information can be correlated with its polarisation and with experimentally measurable lepton angles.
DYnamis provides the Drell--Yan-specific component of the JETHAD environment for carrying out this programme.
Its interface with POWHEG is designed to combine NLO fixed-order information with NLL high-energy evolution while preserving spin correlations and realistic lepton-level kinematics.
The resulting JETHAD--DYnamis--POWHEG framework targets rapidity and transverse-momentum spectra together with azimuthal harmonics, polarisation fractions, and decay-angle distributions.
These observables offer complementary handles on high-energy radiation and establish $Z$+jet production as a natural precision laboratory for testing BFKL dynamics at the LHC.

\section*{Acknowledgments}
\label{sec:acknowledgments}

We are supported by the Atracci\'on de Talento Grant n. 2022-T1/TIC-24176 (Madrid, Spain).

\vspace{-0.05cm}
\begingroup
\setstretch{0.6}
\bibliographystyle{bibstyle}
\bibliography{bibliography}

\begin{thebibliography}{70}
\expandafter\ifx\csname natexlab\endcsname\relax\def\natexlab#1{#1}\fi
\expandafter\ifx\csname bibnamefont\endcsname\relax
  \def\bibnamefont#1{#1}\fi
\expandafter\ifx\csname bibfnamefont\endcsname\relax
  \def\bibfnamefont#1{#1}\fi
\expandafter\ifx\csname citenamefont\endcsname\relax
  \def\citenamefont#1{#1}\fi
\expandafter\ifx\csname url\endcsname\relax
  \def\url#1{\texttt{#1}}\fi
\expandafter\ifx\csname urlprefix\endcsname\relax\def\urlprefix{URL }\fi
\providecommand{\bibinfo}[2]{#2}
\providecommand{\eprint}[2][]{\url{#2}}

\bibitem[{\citenamefont{Benedikt et~al.}(2025{\natexlab{a}})}]{FCC:2025lpp}
\bibinfo{author}{\bibfnamefont{M.}~\bibnamefont{Benedikt}} \bibnamefont{et~al.} (\bibinfo{collaboration}{FCC}), \bibinfo{journal}{Eur. Phys. J. C} \textbf{\bibinfo{volume}{85}}, \bibinfo{pages}{1468} (\bibinfo{year}{2025}{\natexlab{a}}), \eprint{2505.00272}.

\bibitem[{\citenamefont{Benedikt et~al.}(2025{\natexlab{b}})}]{FCC:2025uan}
\bibinfo{author}{\bibfnamefont{M.}~\bibnamefont{Benedikt}} \bibnamefont{et~al.} (\bibinfo{collaboration}{FCC}), \bibinfo{journal}{Eur. Phys. J. ST} \textbf{\bibinfo{volume}{234}}, \bibinfo{pages}{5713} (\bibinfo{year}{2025}{\natexlab{b}}), \eprint{2505.00274}.

\bibitem[{\citenamefont{Benedikt et~al.}(2025{\natexlab{c}})}]{FCC:2025jtd}
\bibinfo{author}{\bibfnamefont{M.}~\bibnamefont{Benedikt}} \bibnamefont{et~al.} (\bibinfo{collaboration}{FCC}), \bibinfo{journal}{Eur. Phys. J. ST} \textbf{\bibinfo{volume}{234}}, \bibinfo{pages}{5113} (\bibinfo{year}{2025}{\natexlab{c}}), \eprint{2505.00273}.

\bibitem[{\citenamefont{Fadin et~al.}(1975)}]{Fadin:1975cb}
\bibinfo{author}{\bibfnamefont{V.~S.} \bibnamefont{Fadin}} \bibnamefont{et~al.}, \bibinfo{journal}{Phys. Lett. B} \textbf{\bibinfo{volume}{60}}, \bibinfo{pages}{50} (\bibinfo{year}{1975}).

\bibitem[{\citenamefont{Balitsky\mbox{, L. N. Lipatov}}(1978)}]{Balitsky:1978ic}
\bibinfo{author}{\bibfnamefont{I.~I.} \bibnamefont{Balitsky\mbox{, L. N. Lipatov}}}, \bibinfo{journal}{Sov.\ J.\ Nucl.\ Phys.} \textbf{\bibinfo{volume}{28}}, \bibinfo{pages}{822} (\bibinfo{year}{1978}).

\bibitem[{\citenamefont{Bacchetta et~al.}(2020)}]{Bacchetta:2020vty}
\bibinfo{author}{\bibfnamefont{A.}~\bibnamefont{Bacchetta}} \bibnamefont{et~al.}, \bibinfo{journal}{Eur. Phys. J. C} \textbf{\bibinfo{volume}{80}}, \bibinfo{pages}{733} (\bibinfo{year}{2020}), \eprint{2005.02288}.

\bibitem[{\citenamefont{Bacchetta et~al.}(2024)}]{Bacchetta:2024fci}
\bibinfo{author}{\bibfnamefont{A.}~\bibnamefont{Bacchetta}} \bibnamefont{et~al.}, \bibinfo{journal}{Eur. Phys. J. C} \textbf{\bibinfo{volume}{84}}, \bibinfo{pages}{576} (\bibinfo{year}{2024}), \eprint{2402.17556}.

\bibitem[{\citenamefont{Amoroso et~al.}(2022)}]{Amoroso:2022eow}
\bibinfo{author}{\bibfnamefont{S.}~\bibnamefont{Amoroso}} \bibnamefont{et~al.}, \bibinfo{journal}{Acta Phys. Polon. B} \textbf{\bibinfo{volume}{53}}, \bibinfo{pages}{A1} (\bibinfo{year}{2022}), \eprint{2203.13923}.

\bibitem[{\citenamefont{Bolognino et~al.}(2018)}]{Bolognino:2018rhb}
\bibinfo{author}{\bibfnamefont{A.~D.} \bibnamefont{Bolognino}} \bibnamefont{et~al.}, \bibinfo{journal}{Eur. Phys. J.} \textbf{\bibinfo{volume}{C78}}, \bibinfo{pages}{1023} (\bibinfo{year}{2018}), \eprint{1808.02395}.

\bibitem[{\citenamefont{Bolognino et~al.}(2021{\natexlab{a}})}]{Bolognino:2021niq}
\bibinfo{author}{\bibfnamefont{A.~D.} \bibnamefont{Bolognino}} \bibnamefont{et~al.}, \bibinfo{journal}{Eur. Phys. J. C} \textbf{\bibinfo{volume}{81}}, \bibinfo{pages}{846} (\bibinfo{year}{2021}{\natexlab{a}}), \eprint{2107.13415}.

\bibitem[{\citenamefont{Hentschinski et~al.}(2023)}]{Hentschinski:2022xnd}
\bibinfo{author}{\bibfnamefont{M.}~\bibnamefont{Hentschinski}} \bibnamefont{et~al.}, \bibinfo{journal}{Acta Phys. Polon. B} \textbf{\bibinfo{volume}{54}}, \bibinfo{pages}{2} (\bibinfo{year}{2023}), \eprint{2203.08129}.

\bibitem[{\citenamefont{Celiberto et~al.}(2021{\natexlab{a}})}]{Celiberto:2020tmb}
\bibinfo{author}{\bibfnamefont{F.~G.} \bibnamefont{Celiberto}} \bibnamefont{et~al.}, \bibinfo{journal}{Eur. Phys. J. C} \textbf{\bibinfo{volume}{81}}, \bibinfo{pages}{293} (\bibinfo{year}{2021}{\natexlab{a}}), \eprint{2008.00501}.

\bibitem[{\citenamefont{Bolognino et~al.}(2021{\natexlab{b}})}]{Bolognino:2021mrc}
\bibinfo{author}{\bibfnamefont{A.~D.} \bibnamefont{Bolognino}} \bibnamefont{et~al.}, \bibinfo{journal}{Phys. Rev. D} \textbf{\bibinfo{volume}{103}}, \bibinfo{pages}{094004} (\bibinfo{year}{2021}{\natexlab{b}}), \eprint{2103.07396}.

\bibitem[{\citenamefont{Bonvini\mbox{, S. Marzani}}(2018)}]{Bonvini:2018ixe}
\bibinfo{author}{\bibfnamefont{M.}~\bibnamefont{Bonvini\mbox{, S. Marzani}}}, \bibinfo{journal}{Phys. Rev. Lett.} \textbf{\bibinfo{volume}{120}}, \bibinfo{pages}{202003} (\bibinfo{year}{2018}), \eprint{1802.07758}.

\bibitem[{\citenamefont{Silvetti\mbox{, M. Bonvini}}(2023)}]{Silvetti:2022hyc}
\bibinfo{author}{\bibfnamefont{F.}~\bibnamefont{Silvetti\mbox{, M. Bonvini}}}, \bibinfo{journal}{Eur. Phys. J. C} \textbf{\bibinfo{volume}{83}}, \bibinfo{pages}{267} (\bibinfo{year}{2023}), \eprint{2211.10142}.

\bibitem[{\citenamefont{Duclou\'e et~al.}(2013)\citenamefont{Duclou\'e, Szymanowski,  Wallon}}]{Ducloue:2013hia}
\bibinfo{author}{\bibfnamefont{B.}~\bibnamefont{Duclou\'e}}, \bibinfo{author}{\bibfnamefont{L.}~\bibnamefont{Szymanowski}},  \bibinfo{author}{\bibfnamefont{S.}~\bibnamefont{Wallon}}, \bibinfo{journal}{JHEP} \textbf{\bibinfo{volume}{05}}, \bibinfo{pages}{096} (\bibinfo{year}{2013}), \eprint{1302.7012}.

\bibitem[{\citenamefont{Celiberto et~al.}(2015)}]{Celiberto:2015yba}
\bibinfo{author}{\bibfnamefont{F.~G.} \bibnamefont{Celiberto}} \bibnamefont{et~al.}, \bibinfo{journal}{Eur. Phys. J. C} \textbf{\bibinfo{volume}{75}}, \bibinfo{pages}{292} (\bibinfo{year}{2015}), \eprint{1504.08233}.

\bibitem[{\citenamefont{Celiberto et~al.}(2016{\natexlab{a}})}]{Celiberto:2016ygs}
\bibinfo{author}{\bibfnamefont{F.~G.} \bibnamefont{Celiberto}} \bibnamefont{et~al.}, \bibinfo{journal}{Eur. Phys. J. C} \textbf{\bibinfo{volume}{76}}, \bibinfo{pages}{224} (\bibinfo{year}{2016}{\natexlab{a}}), \eprint{1601.07847}.

\bibitem[{\citenamefont{Celiberto}(2017)}]{Celiberto:2017ius}
\bibinfo{author}{\bibfnamefont{F.~G.} \bibnamefont{Celiberto}}, \bibinfo{type}{Phd thesis}, \bibinfo{school}{Calabria U. and INFN-Cosenza} (\bibinfo{year}{2017}), \eprint{1707.04315}.

\bibitem[{\citenamefont{Celiberto et~al.}(2022{\natexlab{a}})}]{Celiberto:2022gji}
\bibinfo{author}{\bibfnamefont{F.~G.} \bibnamefont{Celiberto}} \bibnamefont{et~al.}, \bibinfo{journal}{Phys. Rev. D} \textbf{\bibinfo{volume}{106}}, \bibinfo{pages}{114004} (\bibinfo{year}{2022}{\natexlab{a}}), \eprint{2207.05015}.

\bibitem[{\citenamefont{Celiberto et~al.}(2018{\natexlab{a}})}]{Celiberto:2018muu}
\bibinfo{author}{\bibfnamefont{F.~G.} \bibnamefont{Celiberto}} \bibnamefont{et~al.}, \bibinfo{journal}{Phys. Lett.} \textbf{\bibinfo{volume}{B786}}, \bibinfo{pages}{201} (\bibinfo{year}{2018}{\natexlab{a}}), \eprint{1808.09511}.

\bibitem[{\citenamefont{Golec-Biernat et~al.}(2018)}]{Golec-Biernat:2018kem}
\bibinfo{author}{\bibfnamefont{K.}~\bibnamefont{Golec-Biernat}} \bibnamefont{et~al.}, \bibinfo{journal}{JHEP} \textbf{\bibinfo{volume}{12}}, \bibinfo{pages}{091} (\bibinfo{year}{2018}), \eprint{1811.04361}.

\bibitem[{\citenamefont{Celiberto et~al.}(2016{\natexlab{b}})}]{Celiberto:2016hae}
\bibinfo{author}{\bibfnamefont{F.~G.} \bibnamefont{Celiberto}} \bibnamefont{et~al.}, \bibinfo{journal}{Phys. Rev. D} \textbf{\bibinfo{volume}{94}}, \bibinfo{pages}{034013} (\bibinfo{year}{2016}{\natexlab{b}}), \eprint{1604.08013}.

\bibitem[{\citenamefont{Celiberto}(2023{\natexlab{a}})}]{Celiberto:2022kxx}
\bibinfo{author}{\bibfnamefont{F.~G.} \bibnamefont{Celiberto}}, \bibinfo{journal}{Eur. Phys. J. C} \textbf{\bibinfo{volume}{83}}, \bibinfo{pages}{332} (\bibinfo{year}{2023}{\natexlab{a}}), \eprint{2208.14577}.

\bibitem[{\citenamefont{Celiberto et~al.}(2018{\natexlab{b}})}]{Celiberto:2017nyx}
\bibinfo{author}{\bibfnamefont{F.~G.} \bibnamefont{Celiberto}} \bibnamefont{et~al.}, \bibinfo{journal}{Phys. Lett. B} \textbf{\bibinfo{volume}{777}}, \bibinfo{pages}{141} (\bibinfo{year}{2018}{\natexlab{b}}), \eprint{1709.10032}.

\bibitem[{\citenamefont{Bolognino et~al.}(2019)}]{Bolognino:2019yls}
\bibinfo{author}{\bibfnamefont{A.~D.} \bibnamefont{Bolognino}} \bibnamefont{et~al.}, \bibinfo{journal}{Eur. Phys. J. C} \textbf{\bibinfo{volume}{79}}, \bibinfo{pages}{939} (\bibinfo{year}{2019}), \eprint{1909.03068}.

\bibitem[{\citenamefont{Celiberto et~al.}(2021{\natexlab{b}})}]{Celiberto:2021dzy}
\bibinfo{author}{\bibfnamefont{F.~G.} \bibnamefont{Celiberto}} \bibnamefont{et~al.}, \bibinfo{journal}{Eur. Phys. J. C} \textbf{\bibinfo{volume}{81}}, \bibinfo{pages}{780} (\bibinfo{year}{2021}{\natexlab{b}}), \eprint{2105.06432}.

\bibitem[{\citenamefont{Celiberto et~al.}(2021{\natexlab{c}})}]{Celiberto:2021fdp}
\bibinfo{author}{\bibfnamefont{F.~G.} \bibnamefont{Celiberto}} \bibnamefont{et~al.}, \bibinfo{journal}{Phys. Rev. D} \textbf{\bibinfo{volume}{104}}, \bibinfo{pages}{114007} (\bibinfo{year}{2021}{\natexlab{c}}), \eprint{2109.11875}.

\bibitem[{\citenamefont{Celiberto et~al.}(2022{\natexlab{b}})}]{Celiberto:2022zdg}
\bibinfo{author}{\bibfnamefont{F.~G.} \bibnamefont{Celiberto}} \bibnamefont{et~al.}, \bibinfo{journal}{Phys. Rev. D} \textbf{\bibinfo{volume}{105}}, \bibinfo{pages}{114056} (\bibinfo{year}{2022}{\natexlab{b}}), \eprint{2205.13429}.

\bibitem[{\citenamefont{Celiberto}(2022{\natexlab{a}})}]{Celiberto:2022keu}
\bibinfo{author}{\bibfnamefont{F.~G.} \bibnamefont{Celiberto}}, \bibinfo{journal}{Phys. Lett. B} \textbf{\bibinfo{volume}{835}}, \bibinfo{pages}{137554} (\bibinfo{year}{2022}{\natexlab{a}}), \eprint{2206.09413}.

\bibitem[{\citenamefont{Celiberto}(2024{\natexlab{a}})}]{Celiberto:2024omj}
\bibinfo{author}{\bibfnamefont{F.~G.} \bibnamefont{Celiberto}}, \bibinfo{journal}{Eur. Phys. J. C} \textbf{\bibinfo{volume}{84}}, \bibinfo{pages}{384} (\bibinfo{year}{2024}{\natexlab{a}}), \eprint{2401.01410}.

\bibitem[{\citenamefont{Feng et~al.}(2023)}]{Feng:2022inv}
\bibinfo{author}{\bibfnamefont{J.~L.} \bibnamefont{Feng}} \bibnamefont{et~al.}, \bibinfo{journal}{J. Phys. G} \textbf{\bibinfo{volume}{50}}, \bibinfo{pages}{030501} (\bibinfo{year}{2023}), \eprint{2203.05090}.

\bibitem[{\citenamefont{Celiberto}(2023{\natexlab{b}})}]{Celiberto:2022grc}
\bibinfo{author}{\bibfnamefont{F.~G.} \bibnamefont{Celiberto}}, \bibinfo{journal}{Acta Phys. Polon. Supp.} \textbf{\bibinfo{volume}{16}}, \bibinfo{pages}{41} (\bibinfo{year}{2023}{\natexlab{b}}), \eprint{2211.11780}.

\bibitem[{\citenamefont{Boussarie et~al.}(2018)}]{Boussarie:2017oae}
\bibinfo{author}{\bibfnamefont{R.}~\bibnamefont{Boussarie}} \bibnamefont{et~al.}, \bibinfo{journal}{Phys. Rev. D} \textbf{\bibinfo{volume}{97}}, \bibinfo{pages}{014008} (\bibinfo{year}{2018}), \eprint{1709.01380}.

\bibitem[{\citenamefont{Celiberto\mbox{, M. Fucilla}}(2022)}]{Celiberto:2022dyf}
\bibinfo{author}{\bibfnamefont{F.~G.} \bibnamefont{Celiberto\mbox{, M. Fucilla}}}, \bibinfo{journal}{Eur. Phys. J. C} \textbf{\bibinfo{volume}{82}}, \bibinfo{pages}{929} (\bibinfo{year}{2022}), \eprint{2202.12227}.

\bibitem[{\citenamefont{Celiberto}(2023{\natexlab{c}})}]{Celiberto:2023fzz}
\bibinfo{author}{\bibfnamefont{F.~G.} \bibnamefont{Celiberto}}, \bibinfo{journal}{Universe} \textbf{\bibinfo{volume}{9}}, \bibinfo{pages}{324} (\bibinfo{year}{2023}{\natexlab{c}}), \eprint{2305.14295}.

\bibitem[{\citenamefont{Celiberto}(2025{\natexlab{a}})}]{Celiberto:2025ogy}
\bibinfo{author}{\bibfnamefont{F.~G.} \bibnamefont{Celiberto}}, \bibinfo{journal}{Phys. Rev. D} \textbf{\bibinfo{volume}{112}}, \bibinfo{pages}{074023} (\bibinfo{year}{2025}{\natexlab{a}}), \eprint{2506.00776}.

\bibitem[{\citenamefont{Celiberto}(2026{\natexlab{a}})}]{Celiberto:2026qiz}
\bibinfo{author}{\bibfnamefont{F.~G.} \bibnamefont{Celiberto}} (\bibinfo{year}{2026}{\natexlab{a}}), \eprint{2606.20063}.

\bibitem[{\citenamefont{Celiberto\mbox{, A. Papa}}(2024)}]{Celiberto:2023rzw}
\bibinfo{author}{\bibfnamefont{F.~G.} \bibnamefont{Celiberto\mbox{, A. Papa}}}, \bibinfo{journal}{Phys. Lett. B} \textbf{\bibinfo{volume}{848}}, \bibinfo{pages}{138406} (\bibinfo{year}{2024}), \eprint{2308.00809}.

\bibitem[{\citenamefont{Celiberto et~al.}(2024{\natexlab{a}})\citenamefont{Celiberto, Gatto,  Papa}}]{Celiberto:2024mab}
\bibinfo{author}{\bibfnamefont{F.~G.} \bibnamefont{Celiberto}}, \bibinfo{author}{\bibfnamefont{G.}~\bibnamefont{Gatto}},  \bibinfo{author}{\bibfnamefont{A.}~\bibnamefont{Papa}}, \bibinfo{journal}{Eur. Phys. J. C} \textbf{\bibinfo{volume}{84}}, \bibinfo{pages}{1071} (\bibinfo{year}{2024}{\natexlab{a}}), \eprint{2405.14773}.

\bibitem[{\citenamefont{Celiberto}(2024{\natexlab{b}})}]{Celiberto:2024mrq}
\bibinfo{author}{\bibfnamefont{F.~G.} \bibnamefont{Celiberto}}, \bibinfo{journal}{Symmetry} \textbf{\bibinfo{volume}{16}}, \bibinfo{pages}{550} (\bibinfo{year}{2024}{\natexlab{b}}), \eprint{2403.15639}.

\bibitem[{\citenamefont{Celiberto\mbox{, G. Gatto}}(2025)}]{Celiberto:2024beg}
\bibinfo{author}{\bibfnamefont{F.~G.} \bibnamefont{Celiberto\mbox{, G. Gatto}}}, \bibinfo{journal}{Phys. Rev. D} \textbf{\bibinfo{volume}{111}}, \bibinfo{pages}{034037} (\bibinfo{year}{2025}), \eprint{2412.10549}.

\bibitem[{\citenamefont{Celiberto}(2025{\natexlab{b}})}]{Celiberto:2025dfe}
\bibinfo{author}{\bibfnamefont{F.~G.} \bibnamefont{Celiberto}}, \bibinfo{journal}{Phys. Rev. D} \textbf{\bibinfo{volume}{111}}, \bibinfo{pages}{L111501} (\bibinfo{year}{2025}{\natexlab{b}}), \eprint{2504.03949}.

\bibitem[{\citenamefont{Celiberto}(2025{\natexlab{c}})}]{Celiberto:2025ziy}
\bibinfo{author}{\bibfnamefont{F.~G.} \bibnamefont{Celiberto}}, \bibinfo{journal}{Phys. Rev. D} \textbf{\bibinfo{volume}{112}}, \bibinfo{pages}{074041} (\bibinfo{year}{2025}{\natexlab{c}}), \eprint{2507.09744}.

\bibitem[{\citenamefont{Celiberto}(2026{\natexlab{b}})}]{Celiberto:2026kks}
\bibinfo{author}{\bibfnamefont{F.~G.} \bibnamefont{Celiberto}}, \bibinfo{journal}{Phys. Rev. D} \textbf{\bibinfo{volume}{113}}, \bibinfo{pages}{114037} (\bibinfo{year}{2026}{\natexlab{b}}), \eprint{2604.11646}.

\bibitem[{\citenamefont{Celiberto}(2025{\natexlab{d}})}]{Celiberto:2025ipt}
\bibinfo{author}{\bibfnamefont{F.~G.} \bibnamefont{Celiberto}}, \bibinfo{journal}{Eur. Phys. J. C} \textbf{\bibinfo{volume}{85}}, \bibinfo{pages}{1395} (\bibinfo{year}{2025}{\natexlab{d}}), \eprint{2502.11136}.

\bibitem[{\citenamefont{Celiberto}(2026{\natexlab{c}})}]{Celiberto:2026ooh}
\bibinfo{author}{\bibfnamefont{F.~G.} \bibnamefont{Celiberto}}, \bibinfo{journal}{Particles} \textbf{\bibinfo{volume}{9}}, \bibinfo{pages}{23} (\bibinfo{year}{2026}{\natexlab{c}}), \eprint{2604.13769}.

\bibitem[{\citenamefont{Celiberto\mbox{, A. Papa}}(2023)}]{Celiberto:2023rtu}
\bibinfo{author}{\bibfnamefont{F.~G.} \bibnamefont{Celiberto\mbox{, A. Papa}}} (\bibinfo{year}{2023}), \eprint{2305.00962}.

\bibitem[{\citenamefont{Chen et~al.}(2015)}]{Chen:2014gva}
\bibinfo{author}{\bibfnamefont{X.}~\bibnamefont{Chen}} \bibnamefont{et~al.}, \bibinfo{journal}{Phys. Lett. B} \textbf{\bibinfo{volume}{740}}, \bibinfo{pages}{147} (\bibinfo{year}{2015}), \eprint{1408.5325}.

\bibitem[{\citenamefont{Boughezal et~al.}(2015)}]{Boughezal:2015dra}
\bibinfo{author}{\bibfnamefont{R.}~\bibnamefont{Boughezal}} \bibnamefont{et~al.}, \bibinfo{journal}{Phys. Rev. Lett.} \textbf{\bibinfo{volume}{115}}, \bibinfo{pages}{082003} (\bibinfo{year}{2015}), \eprint{1504.07922}.

\bibitem[{\citenamefont{Dawson et~al.}(2022)}]{Dawson:2022zbb}
\bibinfo{author}{\bibfnamefont{S.}~\bibnamefont{Dawson}} \bibnamefont{et~al.} (\bibinfo{year}{2022}), \eprint{2209.07510}.

\bibitem[{\citenamefont{Monni et~al.}(2020)\citenamefont{Monni, Rottoli,  Torrielli}}]{Monni:2019yyr}
\bibinfo{author}{\bibfnamefont{P.~F.} \bibnamefont{Monni}}, \bibinfo{author}{\bibfnamefont{L.}~\bibnamefont{Rottoli}},  \bibinfo{author}{\bibfnamefont{P.}~\bibnamefont{Torrielli}}, \bibinfo{journal}{Phys. Rev. Lett.} \textbf{\bibinfo{volume}{124}}, \bibinfo{pages}{252001} (\bibinfo{year}{2020}), \eprint{1909.04704}.

\bibitem[{\citenamefont{Del~Duca\mbox{, G. Falcioni}}(2025)}]{DelDuca:2025vux}
\bibinfo{author}{\bibfnamefont{V.}~\bibnamefont{Del~Duca\mbox{, G. Falcioni}}}, \bibinfo{journal}{JHEP} \textbf{\bibinfo{volume}{07}}, \bibinfo{pages}{018} (\bibinfo{year}{2025}), \eprint{2504.06184}.

\bibitem[{\citenamefont{Hentschinski et~al.}(2021)}]{Hentschinski:2020tbi}
\bibinfo{author}{\bibfnamefont{M.}~\bibnamefont{Hentschinski}} \bibnamefont{et~al.}, \bibinfo{journal}{Eur. Phys. J. C} \textbf{\bibinfo{volume}{81}}, \bibinfo{pages}{112} (\bibinfo{year}{2021}), \eprint{2011.03193}.

\bibitem[{\citenamefont{Celiberto et~al.}(2022{\natexlab{c}})}]{Celiberto:2022fgx}
\bibinfo{author}{\bibfnamefont{F.~G.} \bibnamefont{Celiberto}} \bibnamefont{et~al.}, \bibinfo{journal}{JHEP} \textbf{\bibinfo{volume}{08}}, \bibinfo{pages}{092} (\bibinfo{year}{2022}{\natexlab{c}}), \eprint{2205.02681}.

\bibitem[{\citenamefont{Celiberto et~al.}(2025{\natexlab{a}})}]{Celiberto:2024bfu}
\bibinfo{author}{\bibfnamefont{F.~G.} \bibnamefont{Celiberto}} \bibnamefont{et~al.}, \bibinfo{journal}{PoS} \textbf{\bibinfo{volume}{DIS2024}}, \bibinfo{pages}{101} (\bibinfo{year}{2025}{\natexlab{a}}), \eprint{2408.08731}.

\bibitem[{\citenamefont{Celiberto\mbox{, F. Lonigro}}(in preparation)}]{Celiberto:2026_Higgs-hadron_NLL-NLO}
\bibinfo{author}{\bibfnamefont{F.~G.} \bibnamefont{Celiberto\mbox{, F. Lonigro}}} (\bibinfo{year}{in preparation}).

\bibitem[{\citenamefont{Celiberto}(2021)}]{Celiberto:2020wpk}
\bibinfo{author}{\bibfnamefont{F.~G.} \bibnamefont{Celiberto}}, \bibinfo{journal}{Eur. Phys. J. C} \textbf{\bibinfo{volume}{81}}, \bibinfo{pages}{691} (\bibinfo{year}{2021}), \eprint{2008.07378}.

\bibitem[{\citenamefont{Celiberto}(2022{\natexlab{b}})}]{Celiberto:2022rfj}
\bibinfo{author}{\bibfnamefont{F.~G.} \bibnamefont{Celiberto}}, \bibinfo{journal}{Phys. Rev. D} \textbf{\bibinfo{volume}{105}}, \bibinfo{pages}{114008} (\bibinfo{year}{2022}{\natexlab{b}}), \eprint{2204.06497}.

\bibitem[{\citenamefont{Celiberto}(2024{\natexlab{c}})}]{Celiberto:2024swu}
\bibinfo{author}{\bibfnamefont{F.~G.} \bibnamefont{Celiberto}}, \bibinfo{journal}{Particles} \textbf{\bibinfo{volume}{7}}, \bibinfo{pages}{502} (\bibinfo{year}{2024}{\natexlab{c}}), \eprint{2405.09526}.

\bibitem[{\citenamefont{Celiberto et~al.}(2023)}]{Celiberto:2023uuk}
\bibinfo{author}{\bibfnamefont{F.~G.} \bibnamefont{Celiberto}} \bibnamefont{et~al.}, in \emph{\bibinfo{booktitle}{{Moriond QCD 2023}}} (\bibinfo{year}{2023}), \eprint{2305.05052}.

\bibitem[{\citenamefont{Celiberto et~al.}(2024{\natexlab{b}})}]{Celiberto:2023eba}
\bibinfo{author}{\bibfnamefont{F.~G.} \bibnamefont{Celiberto}} \bibnamefont{et~al.}, \bibinfo{journal}{PoS} \textbf{\bibinfo{volume}{RADCOR2023}}, \bibinfo{pages}{069} (\bibinfo{year}{2024}{\natexlab{b}}), \eprint{2309.11573}.

\bibitem[{\citenamefont{Celiberto et~al.}(2024{\natexlab{c}})}]{Celiberto:2023nym}
\bibinfo{author}{\bibfnamefont{F.~G.} \bibnamefont{Celiberto}} \bibnamefont{et~al.}, \bibinfo{journal}{PoS} \textbf{\bibinfo{volume}{EPS-HEP2023}}, \bibinfo{pages}{390} (\bibinfo{year}{2024}{\natexlab{c}}), \eprint{2310.16967}.

\bibitem[{\citenamefont{Celiberto et~al.}(2025{\natexlab{b}})}]{Celiberto:2024mdt}
\bibinfo{author}{\bibfnamefont{F.~G.} \bibnamefont{Celiberto}} \bibnamefont{et~al.}, \bibinfo{journal}{PoS} \textbf{\bibinfo{volume}{DIS2024}}, \bibinfo{pages}{126} (\bibinfo{year}{2025}{\natexlab{b}}), \eprint{2408.08757}.

\bibitem[{\citenamefont{Hamilton et~al.}(2013)}]{Hamilton:2012rf}
\bibinfo{author}{\bibfnamefont{K.}~\bibnamefont{Hamilton}} \bibnamefont{et~al.}, \bibinfo{journal}{JHEP} \textbf{\bibinfo{volume}{05}}, \bibinfo{pages}{082} (\bibinfo{year}{2013}), \eprint{1212.4504}.

\bibitem[{\citenamefont{Bagnaschi et~al.}(2023)}]{Bagnaschi:2023rbx}
\bibinfo{author}{\bibfnamefont{E.}~\bibnamefont{Bagnaschi}} \bibnamefont{et~al.}, \bibinfo{journal}{Eur. Phys. J. C} \textbf{\bibinfo{volume}{83}}, \bibinfo{pages}{1054} (\bibinfo{year}{2023}), \eprint{2309.10525}.

\bibitem[{\citenamefont{Banfi et~al.}(2024)}]{Banfi:2023mhz}
\bibinfo{author}{\bibfnamefont{A.}~\bibnamefont{Banfi}} \bibnamefont{et~al.}, \bibinfo{journal}{JHEP} \textbf{\bibinfo{volume}{02}}, \bibinfo{pages}{023} (\bibinfo{year}{2024}), \eprint{2309.02127}.

\bibitem[{\citenamefont{Bonciani et~al.}(2003)}]{Bonciani:2003nt}
\bibinfo{author}{\bibfnamefont{R.}~\bibnamefont{Bonciani}} \bibnamefont{et~al.}, \bibinfo{journal}{Phys. Lett. B} \textbf{\bibinfo{volume}{575}}, \bibinfo{pages}{268} (\bibinfo{year}{2003}), \eprint{hep-ph/0307035}.

\bibitem[{\citenamefont{de~Florian et~al.}(2006)\citenamefont{de~Florian, Kulesza,  Vogelsang}}]{deFlorian:2005fzc}
\bibinfo{author}{\bibfnamefont{D.}~\bibnamefont{de~Florian}}, \bibinfo{author}{\bibfnamefont{A.}~\bibnamefont{Kulesza}},  \bibinfo{author}{\bibfnamefont{W.}~\bibnamefont{Vogelsang}}, \bibinfo{journal}{JHEP} \textbf{\bibinfo{volume}{02}}, \bibinfo{pages}{047} (\bibinfo{year}{2006}), \eprint{hep-ph/0511205}.

\bibitem[{\citenamefont{Muselli et~al.}(2017)\citenamefont{Muselli, Forte,  Ridolfi}}]{Muselli:2017bad}
\bibinfo{author}{\bibfnamefont{C.}~\bibnamefont{Muselli}}, \bibinfo{author}{\bibfnamefont{S.}~\bibnamefont{Forte}},  \bibinfo{author}{\bibfnamefont{G.}~\bibnamefont{Ridolfi}}, \bibinfo{journal}{JHEP} \textbf{\bibinfo{volume}{03}}, \bibinfo{pages}{106} (\bibinfo{year}{2017}), \eprint{1701.01464}.

\end{thebibliography}
\endgroup

\end{document}